\documentclass[aps,prb,twocolumn,groupedaddress,amsmath,amssymb]{revtex4-1}
\usepackage{bm}
\usepackage{graphicx}
\usepackage{siunitx}
\begin{document}
\title{An exactly solvable model of calorimetric measurements}
\author{Brecht Donvil}
\affiliation{Department of Mathematics and Statistics, University of Helsinki, P.O. Box 68, 00014 Helsinki, Finland}
\author{Dmitry Golubev}
\affiliation{Pico group, QTF Centre of Excellence, Department of Applied Physics, Aalto University, P.O. Box 15100, FI-00076 Aalto, Finland}
\author{Paolo Muratore-Ginanneschi}
\affiliation{Department of Mathematics and Statistics, University of Helsinki, P.O. Box 68, 00014 Helsinki, Finland}

\date{\today}
\begin{abstract}
Calorimetric measurements are experimentally realizable methods to assess thermodynamics relations in quantum devices. 
With this motivation in mind, we consider a resonant level coupled to a Fermion reservoir.  
We consider a transient process, in which the interaction between the level and the reservoir is initially switched on and then switched off again.
We find the time dependence of the energy of the reservoir, of the energy of the level and of the interaction energy between them at weak, intermediate, strong and ultra-strong coupling.
We also determine the statistical distributions of these energies.   
\end{abstract}
\maketitle

\section{Introduction}

Large, yet finite reservoirs can simultaneously serve as an environment and a measuring device of the system they are in contact with. 
Indeed, the energy of the reservoir is influenced by interacting with the system and, by measuring it, one can probe the system\cite{Pekola1}.
This idea has motivated a series of recent theory papers \cite{Suomela1,Donvil1,Pekola3,Donvil2,Pekola4}, investigating
temperature variations in a small metallic particle, caused by photon exchange with a qubit.
Here we extend these ideas to an exactly solvable model of a resonant level coupled to a finite size metallic reservoir,
which is the simplest of well-studied quantum impurity models \cite{Bulla}.  
In this model, the variations of the reservoir energy occur due to electron jumps between it and the resonant level.
 
The problem of energy exchange between a microscopic system and the environment  has been studied for a long time
with the main emphasis on the heat to work conversion\cite{Quan,Lutz,Benenti}  and the fluctuation 
relations\cite{Esposito,Campisi,Seifert,Collin,Schuler,Serra}. 
Here we take a different perspective mainly focusing on the variations of the reservoir energy.
In fact, we obtain the time dependent distributions of the reservoir energy, of  
the energy of the resonant level and of the interaction energy. We encompass in our analysis the strong coupling limit,
which favors stronger response of the reservoir energy to the changes in the state of the microscopic system.
We consider transient behavior of the system assuming that the resonant level and the reservoir become coupled at time $t=0$
and, afterwards, the system relaxes to the steady state in the limit $t\to\infty$.
Some of our predictions may be tested in experiments with quantum dots,
which are  well described by the resonant level model.

In related works, the weak coupling regime is typically considered, in which the quantum system is described by a Markovian Lindblad equation\cite{Kosloff}. For example, the distribution of energy emitted into the
environment by a driven two level system has been previously determined in this regime\cite{Gasparinetti,Wollfarth}.
More recently, the strong coupling non-Markovian regime has drawn considerable attention. 
An efficient approximate method of studying strong system-reservoir coupling 
is the reaction coordinate formalism\cite{garg,toss,nazir,strasberg,newman,strasberg2},
in which the reservoir is replaced by a single degree of freedom weakly coupled to the residual environment. 
This method ensures quick convergence to the exact result if one increases the number of reaction coordinates\cite{Martinazzo,Woods}.
The exact strong coupling dynamics of the energy exchange between a driven two level system and its environment 
has been studied within the spin-Boson model\cite{Carrega,Cangemi}. It was found that the strong coupling manifests itself
in the significant role of the interaction energy in the overall energy balance\cite{Carrega,Ankerhold}
and in the non-exponential time relaxation of the energies.

The advantage of the resonant level model, which we discuss here, is that it allows one to study the strong coupling effects
in the energy exchange between the system and the environment exactly and, in the wide band limit, analytically
even in the non-stationary case (see e.g. Ref. \cite{Komnik}). 
This simple model has been studied for a long time and the exact solution of it is presented, for example, in Refs. \cite{cohen,FaPa98,jaksic}.
It has also been proven to be very useful in the context of quantum thermodynamics. 
It has been used, for example, in order to properly define the thermodynamic quatitities of a slowly driven system 
in the strong coupling limit\cite{Ludovico,Esposito2,Bruch,Esposito3}, where the interaction energy cannot be neglected 
and, therefore, it becomes
difficult to separate the heat from the work done on the system. These definitions have been verified numerically in Ref. \cite{Oz},
where the fast driving regime has also been studied. Equilibrium fluctuations of the energy of the resonant level have been 
analyzed in Ref.\cite{Ochoa}. In addition, the resonant level model becomes equivalent to the so called Toulouse limit of the spin-Boson model 
if the level is aligned with the Fermi energy of the reservoir\cite{Guinea}. Finally, a comprehensive introduction on fluctuations of thermodynamic quantities in similar models can be found in Ref\cite{jaksic2}.

The paper is organized as follows: in Sec. \ref{sec:model} we introduce the model
and provide its formal solution in a general form, in Sec. \ref{sec:metal} we apply 
these results to the specific case of metallic reservoir with energy independent 
spectral density of the environment, in Sec. \ref{sec:discussion} we discuss possible
experimental setup, in which our predictions can be tested, and summarize our findings.

\section{Model}
\label{sec:model}

We consider a spin-polarized resonant energy-level coupled to a finite size reservoir. The total system is described by the Hamiltonian
\begin{eqnarray}
\hat H = \hat H_0 + \hat H_R + \hat H_I.
\label{hat_H}
\end{eqnarray}
Here
\begin{equation}
\hat H_0 = \epsilon_0\,\hat c^\dagger_0 \hat c_0
\nonumber
\end{equation}
is the Hamiltonian of the resonant level,
\begin{equation}
\hat H_R=\sum_{k=1}^{\infty}\epsilon_k \hat c^\dagger_k \hat c_k
\nonumber
\end{equation}
is the reservoir Hamiltonian and  
\begin{equation}
%\label{eq:hamiltonianInt}
\hat H_I=\sum_{k=1}^{\infty} g_k(\hat c^\dagger_0 \hat c_k+ \hat c^\dagger_k \hat c_0)
\nonumber
\end{equation}
is the interaction between them.
In addition to the second quantized Hamiltonian (\ref{hat_H}), 
we also define the single particle Hamiltonian of the system 
\begin{eqnarray}
H = H_0 +  H_R +  H_I,
\label{eq:Hh}
\end{eqnarray}
where the Hamiltonians $H_0$,  $H_R$, and $H_I$ are the infinite size self-adjoint matrices of the form 
\begin{eqnarray}
H_0 = \left(\begin{array}{cc}  \epsilon_0 & {\bm 0}^\top \\ {\bm 0} & \tilde 0  \end{array}\right),\;
H_R = \left(\begin{array}{cc}  0 & {\bm 0}^\top \\ {\bm 0} & \tilde\epsilon_R  \end{array}\right), \;
H_I = \left(\begin{array}{cc}  0 & {\bm g}^\top \\ {\bm g} & \tilde 0  \end{array}\right).
\nonumber\\
\label{Hsp}
\end{eqnarray}
Here $\top$ stands for transposition, and ${\bm g}^\top = (g_1,g_2,\dots,g_N)$ is the row vector containing the coupling constants. All matrix elements of the matrix $\tilde 0$ and of the vector ${\bm 0}$ are equal to zero,
and $\tilde\epsilon_R$ is the diagonal matrix containing the energies of the reservoir levels, $(\tilde\epsilon_{R})_{ij}=\epsilon_i\delta_{ij}$.
The Hamiltonian $\hat H$ (\ref{hat_H}) can be expressed as
\begin{eqnarray}
\hat H = \hat{\bm c}^{\dagger} H \hat{\bm c},
\nonumber
\end{eqnarray}
where $\hat{\bm c}^{\dagger}=(\hat {c}_0^{\dagger},\hat {c}_1^{\dagger},\hat {c}_2^{\dagger},\dots)$.

Since the Hamiltonian (\ref{hat_H}) is quadratic, the Fermions do not interact, and one can infer full information
about the energies from the single particle density matrix of the system $\rho$, which has the matrix elements
\begin{eqnarray}
\rho_{m n}(t) =\,{\rm tr} [ \hat c_n^\dagger \hat c_m \hat\rho(t) ].
\nonumber
\end{eqnarray}
Here $\hat\rho(t)$ is the full density matrix of the system in the Fock space.
The density matrix $\rho$ satisfies the usual Liouville-von Neumann equation
\begin{eqnarray}
\imath\,{\partial\rho}/{\partial t} = [H,\rho],
%\label{Sch}
\nonumber
\end{eqnarray}
with solution $$\rho(t) = e^{-\imath\,H\,t}\rho(0) e^{\imath\,H\,t}.$$ Here $\rho(0)$ is an arbitrary initial density matrix. We restrict the attention to the physically relevant case of
a diagonal initial density matrix
\begin{eqnarray}
\rho_{n \,k}(0) = n_{k} \delta_{n\,k}
\nonumber
\end{eqnarray}
with populations specified by a Fermi-Dirac thermal equilibrium distribution
\begin{align}
  n_{k}\equiv n(\epsilon_{k}) =   \frac{1}{1+e^{(\epsilon_k-\mu)/T}}. 
\label{FD}
\end{align}
Here $\mu$ is the chemical potential of the metallic reservoir and $T$ is its initial temperature (Boltzmann constant $k_B=1$).

Due to the simplicity of the model, we can derive an explicit expression for the single particle evolution operator using the Laplace transform,
\begin{eqnarray}
e^{-\imath \,H\,t} = \left(\begin{array}{cc} \varphi & {\bm f}^\top \\ {\bm f} & \tilde F  \end{array}\right).
\label{U}
\end{eqnarray}
Here $ \varphi $ is the occupation amplitude of the resonant level, which is specified by the anti-Laplace transform
\begin{eqnarray}
\varphi(t) = \int_{\mathcal{B} }\frac{\mathrm{d}z}{2\pi}\frac{i e^{-\imath\, z\,t}}{z-\epsilon_0 -\Sigma(z)},
\label{phi}
\end{eqnarray}
on the (rotated) Bromwich contour $$ \mathcal{B}=\left\{ z\,|\,0\,<\,\operatorname{Im}z=\mbox{constant}\ \right\} $$
$\Sigma(z)$ is the self-energy defined for any $ \operatorname{Im}z\,\neq\,0 $ by the
integral
\begin{eqnarray}
\Sigma(z) = \int_{\mathbb{R}}\frac{\mathrm{d}\epsilon}{2\pi}\frac{J(\epsilon)}{z-\epsilon},
\label{Sigma}
\end{eqnarray}
and 
\begin{eqnarray}
J(\epsilon) = 2\pi\sum_{k=1}^\infty g_k^2\,\delta(\epsilon-\epsilon_k)
\nonumber
\end{eqnarray} 
is the environment spectral density. Besides that, we have also introduced the vector ${\bm f}$ with the elements
\begin{eqnarray}
f_k(t) = -\imath\,g_k\int_0^t \mathrm{d}t' e^{-\imath\,\epsilon_k\,(t-t')}\varphi(t'),
%\label{fk}
\nonumber
\end{eqnarray} 
and the square sub-block $\tilde F$ with the matrix elements
\begin{eqnarray}
\tilde F_{n k}(t) = \delta_{nk} e^{-\imath\,\epsilon_k\,t} + \frac{f_n(t) g_k - g_n f_k(t)}{\epsilon_n-\epsilon_k}.
\label{Fij}
\end{eqnarray}
The diagonal matrix elements $\tilde F_{kk}$ should be obtained from (\ref{Fij}) by taking the limit $\epsilon_n\to \epsilon_k$.

Having found the the single particle evolution operator (\ref{U}), we can find the average energy of the resonant level,
\begin{eqnarray}
\langle E_0(t)\rangle =  {\rm tr}\left[H_0\,e^{-\imath \,H\,t}\,\rho(0)\,e^{\imath\,H\,t}\right] = \epsilon_0\rho_{00}(t),
\label{E0}
\end{eqnarray}
where the average occupation probability of the level is
\begin{eqnarray}
\rho_{00}(t) = |\varphi(t)|^2 n_0 + \sum_{k=1}^\infty |f_k(t)|^2n_k.
\label{n0t}
\end{eqnarray}

In the same way, the average change of the reservoir energy can be expressed in the form
\begin{eqnarray}
\langle\Delta E_R(t) \rangle&=&
{\rm tr}\left[H_R\left(e^{-\imath\,H\,t}\,\rho(0)\,e^{\imath\,H\,t}-\rho(0)\right)\right]
\nonumber\\
&= &\sum_{k=1}^\infty \epsilon_k[\rho_{k k}(t) - n_k].
\label{dE1}
\end{eqnarray}
The matrix elements $\rho_{k k}(t)$ can be found from Eq. (\ref{U}),
\begin{eqnarray}
\rho_{k k}(t) = |f_k(t)|^2 n_0 + \sum_{p=1}^\infty |F_{k p}(t)|^2 n_p.
\label{rho_kk}
\end{eqnarray}
Invoking the unitarity of the evolution operator (\ref{U}) we transform Eq. (\ref{dE1}) to an alternative form
\begin{eqnarray}
\langle\Delta E_R\rangle& =& \sum_{k=1}^\infty \epsilon_k\,|f_k|^2\,( n_0 - n_k) 
\nonumber\\
&+& \sum_{n,k=1}^\infty |F_{n\,k}|^2 (\epsilon_n - \epsilon_k)\, n_k,
\label{dER0}
\end{eqnarray} 
which is preferable for a macroscopic reservoir with the dense spectrum because it is insensitive
to the singular behaviour of the matrix elements (\ref{Fij}) at $n=k$. 

The average interaction energy can be inferred from energy conservation
\begin{eqnarray}
\langle E_0(t) \rangle + \langle E_R(t) \rangle + \langle E_I(t) \rangle  = \epsilon_0 n_0 + \sum_{k=1}^\infty \epsilon_kn_k,
\label{conservation}
\end{eqnarray}
and Eqs. (\ref{E0},\ref{dER0}).

Employing the standard methods of full counting statistics for fermions 
\cite{Levitov2,Klich},
one can derive the statistical distributions of the energies $E_R$, $E_0$ and $E_I$, 
\begin{align}
&P_\alpha(E,t)=\nonumber\\&\quad\int\frac{\mathrm{d}\nu}{2\pi} e^{\imath\,E\,\nu}\det\left[ 1-\rho(0) +e^{iHt} e^{-iH_\alpha\nu} e^{-\imath\,H\,t}\rho(0) \right].
\nonumber\\
\label{Pt}
\end{align}  
Here the index $\alpha$ can take the values $\alpha=0,R,I$, and the single particle Hamiltonians $H_\alpha$ are defined in Eq. (\ref{Hsp}).

The simple form of the Hamiltonian $H_0$ allows one to find the distribution of the energy of the resonant level exactly,
\begin{eqnarray}
P_0(E,t) = [1-\rho_{00}(t)]\delta(E) + \rho_{00}(t)\delta(E-\epsilon_0).
\label{P0} 
\end{eqnarray}
The result (\ref{P0})  implies that the energy of the resonant level
randomly jumps between the two fixed values --- $\epsilon_0$,  corresponding to the occupied level, and 0, corresponding to the empty level.

One can also derive an explicit expression for the probability distribution of the interaction energy.
The interaction Hamiltonian $H_I$ has three eigenvalues: multiple degenerate eigenvalue 0 and two non-degenerate eigenvalues with the opposite signs $\pm\Delta E_I$, where
\begin{eqnarray}
\Delta E_I=\sqrt{\sum_{k=1}^\infty g_k^2} 
\label{dEI}
\end{eqnarray}
is the "quantum" of the interaction energy.
The corresponding eigenvectors have the components 
$$|\psi_\pm\rangle^\top = (1/\sqrt{2})(1, \pm {\bm g}^\top/\Delta E_I).$$
Re-writing the determinant in the Eq. (\ref{Pt}) in the basis of the eigenvectors of the matrix $H_I$, one observes that it
reduces to the determinant of a simple $2\times 2$ matrix. The latter can be evaluated, which results in the following energy distribution
\begin{eqnarray}
&& P_I(E,t) =  [1-W_+(t) - W_-(t)]\delta(E) 
\nonumber\\ &&
+\, W_+(t)\delta(E-\Delta E_I) + W_-(t)\delta(E+\Delta E_I).
\label{PI}
\end{eqnarray}
Here the probabilities $W_\pm(t)$ have the form 
\begin{eqnarray}
W_+(t) &=& \rho_{++}(1 - \rho_{--}) +  |\rho_{+-}|^2,
\nonumber\\
W_-(t) &=& \rho_{--}(1 - \rho_{++}) +  |\rho_{+-}|^2,
\nonumber
\end{eqnarray}
and the matrix elements of the density matrix are defined as $$\rho_{s s'}(t)=\langle\psi_s| e^{-\imath\,H\,t}\rho(0) e^{\imath\,H\,t}|\psi_{s'}\rangle ,$$ and
$s,s'=\pm$.

The distribution of the reservoir energy cannot be found exactly.
However, its general form in the weak coupling limit can be easily figured out.
Indeed, in this case one can ignore the interaction energy $E_I$ and apply the
energy conservation condition (\ref{conservation}) for instantaneous, fluctuating,
values of the energies $E_0$ and $\Delta E_R$.
This approximation corresponds to the quantum jump approach often used in quantum optics. 
From the Eq. (\ref{conservation}) one then obtains $\Delta E_R(t)=\epsilon_0n_0-E_0(t)$,
and the distribution of the reservoir energy follows from the Eq. (\ref{P0}),
\begin{eqnarray}
P_R(\Delta E,t) &=& [1-\rho_{00}(t)]\delta(\Delta E-\epsilon_0n_0) 
\nonumber\\ &&
+\, \rho_{00}(t)\delta(\Delta E-\epsilon_0(n_0-1)).
\label{PER} 
\end{eqnarray}
If the energy level is initially occupied, $n_0=1$, the distribution (\ref{PER}) has
one peak at $\Delta E_R=0$ and the second peak -- at positive energy $\Delta E_R=\epsilon_0$.
In this case, the reservoir energy can either stay unchanged or increase by $\epsilon_0$ if an electron leaves 
the level $\epsilon_0$ and enters the reservoir.
If the energy level is initially empty, $n_0=0$, the peaks of the distribution (\ref{PER})
occur at $\Delta E_R=0$ and $\Delta E_R=-\epsilon_0$. The latter peak describes the reduction
of the reservoir energy, which occurs if an electron leaves the reservoir and populates the level $\epsilon_0$. 
If one goes beyond the weak coupling limit, the $\delta-$peaks in the distribution (\ref{PER}) acquire
finite width $\sim |{\rm Im}(\Sigma(\epsilon_0))|$. %Here we study this regime only numerically.

\section{Metallic reservoir}
\label{sec:metal}

We now apply the general results presented in the previous section to an important example: a metallic reservoir. We thus consider a spectral density non-vanishing and constant
\begin{eqnarray}
J(\epsilon) = \Gamma_0 \,\theta(\epsilon_c-\epsilon)\,\theta(\epsilon)
\nonumber
\end{eqnarray}
in a region delimited by sharp cut-offs at $\epsilon=0  $ and at $ \epsilon=\epsilon_{c}\,>\,0 $. 
The precise value of the cutoff energy $\epsilon_c$ is not important because most of the measurable parameters depend on it
logarithmically.

\subsection{Long time asymptotics, general analysis}
\label{long_time}

\begin{figure}
\includegraphics[scale=1]{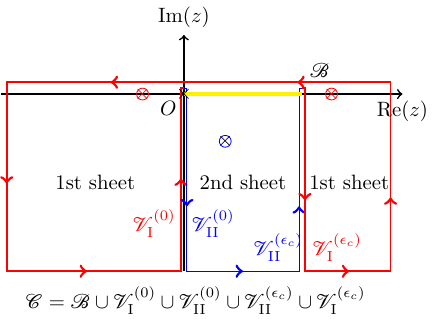}
	\caption{The contour $ \mathcal{C} $ incorporating the Bromwich contour $ \mathcal{B} $ of the Laplace anti-transform (\ref{phi}) for positive $t$. $ \mathcal{B} $, $ \mathcal{V}^{(0)}_{\mathrm{I}} $ and $ \mathcal{V}^{(\epsilon_{c})}_{\mathrm{I}} $ (red online) are the contour components lying on the first Riemann sheet.  $ \mathcal{V}^{(0)}_{\mathrm{II}} $ and $ \mathcal{V}^{(\epsilon_{c})}_{\mathrm{II}} $ (blue online)  are on the second Riemann sheets. The remaining parts of the contour are at infinity and give vanishing contributions to the occupation amplitude. Circled $\times $ stand for poles encompassed by the contour	\label{Fig:contour}}
\end{figure}

The  self-energy (\ref{Sigma}) of the metalic reservoir model  becomes
\begin{align}
\Sigma(z)=-\frac{\Gamma_{0}}{2\,\pi} \big{(}\ln(z-\epsilon_{c})-\ln(z)\big{)}
\nonumber
\end{align}
for any complex $ z $  outside a branch cut in $ z\in [0,\epsilon_{c}] $. When crossing the branch cut from positive to negative values of  $ \operatorname{Im} z $, the self energy develops a discontinuity proportional to the intensity of the spectral density
\begin{align}
\lim_{\varepsilon\downarrow 0}\big{(}\Sigma(x+\imath\epsilon)-\Sigma(x-\imath\epsilon)\big{)}=-\imath\,\Gamma_{0}
\nonumber
\end{align} 
Accordingly, we may analytically extend the self-energy to a second Riemann sheet by requiring continuity across the cut
\begin{align}
\Sigma_{\mathrm{II}}(z)=\Sigma(z)-\imath\,\Gamma_{0}.
\nonumber
\end{align}
In doing so, we also adopt the convention that the principal value $ \operatorname{Arg}(z) $ of any complex number $ z $ 
is in the range $ (-\pi\,,\pi] $. 
These considerations are useful in light of the fact that for any $t>0$ we can most conveniently analyze \cite{FaPa98} the integral specifying $ \varphi $ by embedding the Bromwich path in the closed contour $ \mathcal{C}$ shown in Fig.~\ref{Fig:contour}. The contour treads the first and second Riemann sheets.    
We can then apply Cauchy theorem to write the occupation probability as
\begin{align}
\varphi(t)=\sum_{\wp \in \mathcal{P}}\mathcal{R}_{\wp}(t)+I_{\mathcal{V}_{\mathrm{I}}^{(0)}\cup\mathcal{V}_{\mathrm{II}}^{(0)}}(t)+I_{\mathcal{V}_{\mathrm{I}}^{(\epsilon_{c})}\cup\mathcal{V}_{\mathrm{II}}^{(\epsilon_{c})}}(t)
\label{complex}
\end{align}
The sum ranges over the residues of the poles $ \wp $ enclosed by the contour $ \mathcal{C} $ whereas 
\begin{align}
\begin{array}{l}
I_{\mathcal{V}_{\mathrm{I}}^{(0)}\cup\mathcal{V}_{\mathrm{II}}^{(0)}}(t)
=\int_{-\infty}^{0} \dfrac{\mathrm{d}y}{2\,\pi} \dfrac{-\imath\,\Gamma_{0}\,e^{y\,t}}{D(\imath y) \,D_{\mathrm{II}}(\imath y) }
\\[0.3cm]
I_{\mathcal{V}_{\mathrm{I}}^{(\epsilon_{c})}\cup\mathcal{V}_{\mathrm{II}}^{(\epsilon_{c})}}(t)
=\int_{-\infty}^{0} \dfrac{\mathrm{d}y}{2\pi} \dfrac{\imath\,\Gamma_{0}\,e^{-\imath\,\epsilon_{c}\,t+ y\,t}}{D(\epsilon_{c}+\imath y) \,D_{\mathrm{II}}(\epsilon_{c}+\imath y) }
\end{array}
\label{vertical}
\end{align}
and
\begin{align}
D(z)=z-\epsilon_{0}-\Sigma(z) \,, \hspace{0.5cm}D_{\mathrm{II}}(z)=z-\epsilon_{0}-\Sigma_{\mathrm{II}}(z) 
\nonumber
\end{align}
A detailed study of the analytic properties of the occupation probability integrand \cite{Wolkanowski2013} shows that poles on the first Riemann sheet can only occur on the real line outside the branch cut. Poles on the first Riemann sheet are thus solutions of
\begin{align}
x-\epsilon_{0}+\frac{\Gamma_{0}}{4\,\pi}\ln\frac{(x-\epsilon_{c})^{2}}{x^{2}}=0
\label{poles:I}
\end{align}
and physically bring about Rabi-like oscillations in the occupation amplitude of the resonant level.
Conversely, in the region of the second Riemann sheet enclosed by the contour $ \mathcal{C} $ of Fig~\ref{Fig:contour} poles are solutions of the system
	\begin{align}
	\begin{array}{l}
	 x-\epsilon_{0}+\dfrac{\Gamma_{0}}{4\,\pi}\ln\dfrac{(x-\epsilon_{c})^{2}+y^{2}}{x^{2}+y^{2}}=0
	\\
	y+\dfrac{\Gamma_{0}}{2\,\pi}\left(\arctan\dfrac{y}{x-\epsilon_{c}}-\arctan\dfrac{y}{x}\right)+\dfrac{\Gamma_{0}}{2}=0
	\end{array}
		\label{second:pole}
	\end{align}
for $$ z=x+\imath\,y.$$
In general, residues of  poles  with finite imaginary part lead to  exponentially decaying contributions to a probability amplitude.

Finally, the contributions of the vertical contours (\ref{vertical}) are proportional to $ \Gamma_{0} $ at small coupling, and to $ \Gamma_{0}^{-1} $ at very large coupling.
Furthermore, we observe that for finite $ t $ the  integrands in (\ref{vertical}) differ significantly from zero for energies of the order $ 1/t $.
 Upon expanding the denominators around $ 1/t $ we obtain at leading order  the estimate
\begin{align}
&I_{\mathcal{V}_{\mathrm{I}}^{(0)}\cup\mathcal{V}_{\mathrm{II}}^{(0)}}(t)+I_{\mathcal{V}_{\mathrm{I}}^{(\epsilon_{c})}\cup\mathcal{V}_{\mathrm{II}}^{(\epsilon_{c})}}(t)
\nonumber\\
& \approx \dfrac{\imath\,\Gamma_{0}\,e^{-\imath\,\epsilon_{c}\,t}}{t\,D(\epsilon_{c}+\imath/t) \,D_{\mathrm{II}}(\epsilon_{c}+\imath/t) }
-\dfrac{\imath\,\Gamma_{0}}{t\,D(\imath/t) \,D_{\mathrm{II}}(\imath/t) }
\label{longtime}
\end{align}
The accuracy of the estimate improves as time elapses (see Fig.~\ref{fig:logD}).  For
\begin{align}
t \gg \frac{1}{\epsilon_{c}}e^{\frac{3\,\pi}{2}}\,, \frac{1}{\epsilon_{c}}e^{\frac{2\,\pi\,|\epsilon_{c}-\epsilon_{0}|}{\Gamma_{0}}}
\nonumber
\end{align}
the estimate (\ref{longtime}) reduces to the simpler expression
\begin{align}
&I_{\mathcal{V}_{\mathrm{I}}^{(0)}\cup\mathcal{V}_{\mathrm{II}}^{(0)}}(t)+I_{\mathcal{V}_{\mathrm{I}}^{(\epsilon_{c})}\cup\mathcal{V}_{\mathrm{II}}^{(\epsilon_{c})}}(t)
\nonumber\\
&\overset{t\uparrow \infty}{\to}
\frac{4\,\pi\,e^{-\imath\frac{\epsilon_{c}\,t}{2}}}{\Gamma_{0}\,t\,\ln^{2} (\epsilon_{c}\,t)}\sin\left(\frac{\epsilon_{c}\,t}{2}\right)
\label{longtime2}
\end{align} 
The derivation of further analytic asymptotic expressions hinges upon
the introduction of explicit assumptions on the strength of the coupling.

\subsection{Numerical analysis}\label{sec:num}

We numerically compute the evolution operator \eqref{U} by direct
exponentiation of the single particle Hamiltonian \eqref{eq:Hh} 
for a finite amount of energy levels $N$ in the reservoir.
We keep the level spacing in the reservoir constant $\Delta\omega=\epsilon_c/N$ and take that coupling constants to be
\begin{equation}
g_k=\sqrt{\frac{\Gamma_0\Delta \omega}{2\pi }}.
\nonumber
\end{equation}
Note that we scale the coupling with the level spacing $\Delta\omega$ such that the total coupling does increase with increased reservoir size \cite{Rivas}. Having computed \eqref{U}, we are able to evaluated Eqs. \eqref{E0} and \eqref{dER0} for the energy of the resonant level and the change in reservoir energy.

To obtain the full probability distributions, we take a slightly different route and directly evaluate Eq. \eqref{Pt}. We compute the determinant by numerically exponentiating the matrices. Finally, we perform the Fourier transform using the FFT algorithm.

\subsection{Weak and intermediate coupling asymptotic analysis}

We assume the following relation between the parameters:
\begin{eqnarray}
\Gamma_0, T\ll \mu,\epsilon_0 \ll \epsilon_c.
\label{condition}
\end{eqnarray}
The relation between the $\Gamma_0$, the temperature $T$ and the detuning between the level and the Fermi energy, $|\epsilon_0-\mu|$, can be arbitrary.
The condition (\ref{condition}) covers most physically relevant situations, in which a metallic reservoir with big Fermi energy is involved.
It also includes the regime of intermediate coupling, where the deviations from Markovian dynamics already become significant.

In this regime, residues of the poles on the first Riemann sheet
do not give any sizeable contribution. Although (\ref{poles:I}) admits two roots outside the branch cut for any positive $\Gamma_{0}$, such roots emerge from the branch cut
end points with a non analytic dependence upon $\Gamma_{0}$:
$$ 
x_{-}\sim -\epsilon_{c}\,e^{-2\pi\frac{\epsilon_{0}}{\Gamma_{0}}},
\hspace{0.5cm}
x_{+}\sim \epsilon_{c}+ \epsilon_{c} e^{-2\pi\frac{\epsilon_{c}-\epsilon_{0}}{\Gamma_{0}}}
$$
and the corresponding residues are exponentially suppressed. 

On the second Riemann sheet (\ref{second:pole}) admits within leading order accuracy in the coupling the solution 
\begin{align}
    z_{*}=\tilde{\epsilon}_0+\imath \frac{\Gamma_{0}}{2}
    \label{pole:decay}
\end{align}
The quantity
\begin{eqnarray}
\tilde\epsilon_0 = \epsilon_0 - \frac{\Gamma_0}{2\pi}\ln\left|\frac{\epsilon_c-\epsilon_0}{\epsilon_0}\right|
\nonumber
\end{eqnarray}
physically describes the energy of the resonant level shifted due to its coupling to the reservoir.
%the self-energy (\ref{Sigma}) takes the form
%\begin{eqnarray}
%\Sigma(E) = -\frac{\Gamma_0}{2\pi}\ln\left|\frac{\epsilon_c-E}{E}\right| -i\frac{\Gamma_0}{2}.
%\label{Sigma_m}
%\end{eqnarray} 
%Under the conditions (\ref{condition}) one can replace $E\to \epsilon_0$ under the logarithm. 
%Afterwards, one finds
Upon evaluating the residue of (\ref{pole:decay}) and recalling
that the contributions (\ref{vertical}) are proportional to $ \Gamma_{0} $ at small coupling, 
within leading accuracy we get
\begin{eqnarray}
\begin{array}{l}
 \varphi(t) = e^{-i\tilde\epsilon_0t}e^{-\Gamma_0\,t/2},
\\[0.3cm]
f_k(t) = g_k \dfrac{e^{-i\epsilon_k t} - e^{-i\tilde\epsilon_0 t}e^{-\Gamma_0t/2}}{\epsilon_k-\tilde\epsilon_0+i\frac{\Gamma_0}{2}},
\end{array}
\label{weak:phif}
\end{eqnarray}
The self-consistency condition for (\ref{weak:phif}) is 
$$ t \,\ll\, t_{K}$$ where  $t_{K}$ is the time scale predicted by Khalfin's theorem after which unitary dynamics forbids exponential decay \cite{Khalfin}. We estimate $t_{K}$ by matching exponential
decay with the involution of the longtime asymptotics (\ref{longtime2})
$$ 
e^{-\Gamma_0\,t_{K}/2} \sim \frac{4\,\pi\,}{\Gamma_{0}\,t_{K}\,\ln^2(\epsilon_{c}\,t_{K})}
$$
In particular, if  model parameters  are as in Fig.~\ref{Fig:Et}, we find that  for $ \Gamma_{0}=0.001\,\mu $  Khalfin's time is 
$ \Gamma_0 t_{K} \approx 9.1$ whereas for 
$ \Gamma_{0}=0.02\,\mu $  we get $ \Gamma_0 t_{K}\approx 7.4$. In both cases $e^{-\Gamma_0t_K}\ll 1$, which indicates that
one can use the approximation (\ref{weak:phif}) during the whole relaxation process. 
For small long time tails one should use better a approximation, which was outlined in Sec. \ref{long_time}.

The occupation probability of the level (\ref{n0t}) becomes
\begin{eqnarray}
&& \rho_{00} =  e^{-\Gamma_0 t} n_0 
\nonumber\\ &&
+\, \Gamma_0\int_{0}^{\epsilon_c}\frac{d\epsilon}{2\pi} 
\frac{1+e^{-\Gamma_0t} - 2e^{-\frac{\Gamma_0t}{2}}\cos[(\epsilon-\tilde\epsilon_0)t]}{(\epsilon-\tilde\epsilon_0)^2 + \frac{\Gamma_0^2}{4}} n_F(\epsilon),
\nonumber\\
\label{n0}
\end{eqnarray}
where $n_F(\epsilon)$ is the Fermi function (\ref{FD}) now evaluated over the continuous variable $\epsilon$.
The occupation probabilities of the levels in the reservoir (\ref{rho_kk})
become
\begin{eqnarray}
&& \rho_{kk} =
n_k + \frac{1+e^{-\Gamma_0t} - 2e^{-\frac{\Gamma_0t}{2}}\cos[(\epsilon_k-\tilde\epsilon_0)t]}{(\epsilon_k-\tilde\epsilon_0)^2 + \frac{\Gamma_0^2}{4}}[n_0-n_k]
\nonumber\\ &&
+\,g_k^2\int_0^{\epsilon_c}\frac{d\epsilon}{2\pi}\frac{\Gamma_0[n_F(\epsilon)-n_k]}{(\epsilon-\epsilon_k)^2}
\nonumber\\ &&\times\,
\left| \frac{e^{-i\epsilon t}-e^{-i\tilde\epsilon_0t} e^{-\frac{\Gamma_0t}{2}}}{\epsilon - \tilde\epsilon_0+i\frac{\Gamma_0}{2}} 
- \frac{e^{-i\epsilon_kt}-e^{-i\tilde\epsilon_0t} e^{-\frac{\Gamma_0t}{2}}}{\epsilon_k-\tilde\epsilon_0+i\frac{\Gamma_0}{2}} \right|^2.
\label{nk}
\end{eqnarray}

The average value of the energy of the resonant level, $\langle E_0(t)\rangle$, is given by the Eq. (\ref{E0}) in which
$\rho_{00}(t)$ has the form (\ref{n0}). 
Substituting the expression (\ref{nk}) in Eq. (\ref{dER0}) and replacing the summation over $k$  by the
integral over the energy $\epsilon_k$, we obtain the average change of the reservoir energy in the form
\begin{eqnarray}
&& \frac{\langle\Delta E_R(t)\rangle}{\Gamma_0} =  \int_0^{\epsilon_c}\frac{d\epsilon}{2\pi}
\bigg[ 
\frac{\epsilon  [n_0-n_F(\epsilon)] + \frac{\Gamma_0}{2\pi}\ln\frac{\epsilon_c-\epsilon_0}{\epsilon_0} n_F(\epsilon)}{(\epsilon-\tilde\epsilon_0)^2+\frac{\Gamma_0^2}{4}}
\nonumber\\ && \times\,
\left(1+e^{-\Gamma_0t}-2e^{-\Gamma_0t/2}\cos[(\epsilon-\tilde\epsilon_0)t]\right)
\nonumber\\ &&
-\, \frac{(1-e^{-\Gamma_0t})(\epsilon-\tilde\epsilon_0) - \Gamma_0 e^{-\Gamma_0t/2} \sin[(\epsilon-\tilde\epsilon_0)t]}
{(\epsilon-\tilde\epsilon_0)^2 + \frac{\Gamma_0^2}{4}} n_F(\epsilon)\bigg].
\nonumber\\
\label{dER}
\end{eqnarray}
The average value of the interaction energy $\langle E_I(t)\rangle$ can be inferred from the energy conservation  (\ref{conservation}).

In the weak coupling regime 
\begin{eqnarray}
\Gamma_0 \ll (\tilde\epsilon_0-\mu)\coth\frac{\tilde\epsilon_0-\mu}{2T}
%\label{weak_coupling}
\nonumber
\end{eqnarray}
the integrals (\ref{n0},\ref{dER}) can be straightforwardly evaluated. 
We obtain simple exponential relaxation of the energies, 
which is a typical feature of Markovian Lindblad approximation,
\begin{eqnarray}
\langle E_0(t)\rangle &=& e^{-\Gamma_0 t} \epsilon_0n_0 + (1-e^{-\Gamma_0t}) \epsilon_0 n_F(\epsilon_0), 
\nonumber\\
\langle\Delta E_R(t)\rangle &=& (1-e^{-\Gamma_0t})\epsilon_0[n_0-n_F(\epsilon_0)].
\label{dER_weak}
\end{eqnarray}
The interaction energy in this regime is negligible.

In the long time limit $t_{K}\,\gg \,t\gg (\pi T)^{-1}$ 
the occupation probablities of the levels (\ref{n0}) and (\ref{nk}) approach their asymptotic values
\begin{eqnarray}
&& \rho_{00}^{\textrm{as}} = \frac{1}{2}
-\frac{1}{\pi}{\rm Im}\left[\Psi\left(\frac{1}{2} + \frac{\Gamma_0}{4\pi T} + \imath\,\frac{\tilde\epsilon_0-\mu}{2\,\pi T}\right)\right],
\nonumber\\
&& \rho_{kk}^{\textrm{as}} = n_k + \frac{g_k^2}{(\epsilon_k-\tilde\epsilon_0)^2+\frac{\Gamma_0^2}{4}}
\bigg\{n_0 + \rho_{00}^{\rm as} - 2n_k 
\nonumber\\ &&
+\,  \frac{\Gamma_0}{\pi}{\rm Re}
\bigg[\frac{\Psi\left(\frac{1}{2} + \frac{\Gamma_0}{4\,\pi T} + \imath\,\frac{\tilde\epsilon_0-\mu}{2\,\pi T}\right)
- \Psi\left(\frac{1}{2} + \imath\,\frac{\epsilon_k-\mu}{2\,\pi T}\right)}{\epsilon_k-\tilde\epsilon_0+i\frac{\Gamma_0}{2}}
\bigg]
\nonumber\\ &&
-\, \frac{\Gamma_0}{2\pi^2 T}\,{\rm Im} \bigg[\Psi'\left(\frac{1}{2} + \imath\,\frac{\epsilon_k-\mu}{2\,\pi T}\right)\bigg]
\bigg\},
\label{rho_kk2}
\end{eqnarray}
where $\Psi(x)$ is the digamma function. Accordingly, the energies in the long time limit take the form
\begin{eqnarray}
\langle E_0^{\textrm{as}}\rangle &=& \epsilon_0 \rho_{00}^{\rm as},
\label{E0_eq}\\ 
\langle \Delta E_R^{\textrm{as}} \rangle &=& 
\epsilon_0[n_0-\rho_{00}^{\rm as}] - \langle E_I^{\rm as} \rangle,
\label{ER_eq}\\
\langle E_I^{\textrm{as}} \rangle 
&=& - \frac{\Gamma_0}{\pi}\bigg\{\ln\frac{\epsilon_c-\epsilon_0}{\epsilon_0} \rho_{00}^{\rm as} + \ln\frac{\epsilon_0}{2\pi T}
\nonumber\\ && 
- \,{\rm Re}\left[ \Psi\left(\frac{1}{2} + \frac{\Gamma_0}{4\pi T} + \imath \frac{\tilde\epsilon_0-\mu}{2\,\pi T}\right) \right]\bigg\}.
\label{EI_eq}
\end{eqnarray}
The asymptotic distribution function in the metallic reservoir (\ref{rho_kk2}) has a Lorentzian peak or dip
close to $\tilde{\epsilon}_{0}$. Clearly, such strongly non-equilibrium distribution 
will relax to the thermal one during the electron-electron relaxation time $\tau_{e-e}$. Thus, the distribution (\ref{rho_kk2}) 
survives only during the time interval $(\pi T)^{-1} \,<\, t\, <\,\tau_{e-e}$.
This condition provides the range of validity of our model.

\subsection{Ultra strong coupling asymptotic analysis}
%\label{sec:ultrs}

We now assume that $\Gamma_{0}$ sets the largest energy scale in the model:
\begin{align}
    \epsilon_{0},\epsilon_{c}\,\ll\,\Gamma_{0}
    \nonumber
\end{align}
In this case, the residues of poles in the first and second Riemann sheets exchange their roles in relation to their significance for the dynamics.
Namely, (\ref{poles:I}) admits two solutions
\begin{align}
   & x_{\pm}=\pm \Omega_{1}+ \Omega_{2}+O\left( \Gamma_{0}^{-1}\right)
\label{ultra:real}
\end{align}
with
\begin{align}
&\Omega_{1}=\sqrt{\frac{\Gamma_{0}\,\epsilon_{c}}{2\,\pi}}
   +\sqrt{\frac{2\,\pi}{\Gamma_{0}}}\frac{7\,\epsilon_{c}^{2}+12\,\epsilon_{0}(\epsilon_{0}-\epsilon_{c})}{96\sqrt{\epsilon_{c}}}
   \nonumber \\
%\hspace{0.8cm}
&   \Omega_{2}=\frac{2 \epsilon_{0}+\epsilon_{c}}{4}
    \nonumber
\end{align}
appearing to the right ($x_{+}$) and to the left ($x_{-}$) of the branch cut. These solutions correspond to two simple poles with residues $\mathcal{R}_{\pm}$ now giving an $O(1)$ contribution to (\ref{complex}):
\begin{align}
&\mathcal{R}_{+}+\mathcal{R}_{-}=e^{-\imath \Omega_{2}\,t}\left(1-\frac{\pi\,\epsilon_{c}}{6\,\Gamma_{0}}\right)\,\cos(\Omega_{1}t)
\nonumber\\
&+\imath\,e^{-\imath \Omega_{2}\,t}\sqrt{\pi}\frac{\epsilon_{c}-2\,\epsilon_{0}}{\sqrt{8\,\epsilon_{c}\,\Gamma_{0}}}\sin(\Omega_{1}t)+O(\Gamma_{0}^{-3/2})
\nonumber
\end{align}
Note that the energy $\Omega_1$ approaches the interaction energy quantum (\ref{dEI}) in the limit of infinitely strong coupling, 
$\Omega_1=\Delta E_I$ for $\Gamma_0\to\infty$.
On the second Riemann sheet, (\ref{second:pole}) admits the
solution 
\begin{align}
z_{\star}=\epsilon_{0}+\imath \frac{\Gamma_{0}}{2}+O(\Gamma_{0}^{-1}).
\nonumber
\end{align}
The imaginary part of the root entails an exponential suppression of the residue with very large rate. As a consequence, the contribution to (\ref{complex}) is negligible after an elapse of any non-vanishing time interval $t$. Finally, (\ref{longtime}) estimates the contribution of the the vertical tracts of the complex plane contour $\mathcal{C}$ as of the order $ (\Gamma_{0}\,t)^{-1} $ for finite $ t $.
%Finally, we can estimate contribution of the the vertical tracts of the complex plane contour $\mathcal{C}$ as
%\begin{align}
% I_{\mathcal{V}_{\mathrm{I}}^{(0)}\cup\mathcal{V}_{\mathrm{II}}^{(0)}}(t)+I_{\mathcal{V}_{\mathrm{I}}^{(\epsilon_{c})}\cup\mathcal{V}_{\mathrm{II}}^{(\epsilon_{c})}}(t)\lessapprox 
%\frac{4\,\,e^{-\imath\frac{\epsilon_{c}\,t}{2}}}{\Gamma_{0}\,t\,\pi}\sin\left(\frac{\epsilon_{c}\,t}{2}\right)   
%\nonumber
%\end{align}
The upshot is that energy statistics within leading order accuracy
are dominated by stable oscillations determined by the residues of the first Riemann sheet poles (\ref{ultra:real}). Indeed, 
\begin{eqnarray}
 \varphi(t) &=& e^{-\imath \Omega_{2}\,t}\cos(\Omega_{1}t)+O(\Gamma_{0}^{-1/2})
\label{phi_strong}
\\
f_{k}(t) &=& -\imath g_{k} \dfrac{e^{-\imath\,\Omega_{2}\, t}}{\Omega_{1}}\sin(\Omega_{1}t)\left(1+O(\Gamma_{0}^{-1/2})\right)
\end{eqnarray}
yield (up to corrections $O(\Gamma_{0}^{-1/2})$)
\begin{eqnarray}
\langle E_0(t)\rangle &=& \epsilon_{0} n_0 \cos^{2}(\Delta E_I t)
\nonumber\\ &&
+\,\epsilon_{0}\sin^{2}(\Delta E_I t)
\int_{0}^{\epsilon_c}\frac{d\epsilon}{\epsilon_{c}} 
 n_F(\epsilon)
\label{ultra:n0}
\end{eqnarray}
and
\begin{eqnarray}
\langle\Delta E_R(t) \rangle= \sin^{2}(\Delta E_I t) \int_{0}^{\epsilon_c}\frac{d\epsilon}{\epsilon_{c}} \epsilon\,
 \big{(} n(0)-n(\epsilon) \big{)}.
\label{dER_ultra}
\end{eqnarray}
In this case, the reservoir can effectively be replaced by a single energy level in accordance with the theory
of fermionic reaction coordinates\cite{strasberg2} and results from the spectral analysis of the Hamiltonian operator where at strong coupling one finds that the one particle Hamiltonian gains a pure point spectrum, see e.g. \cite{jaksic,Cornean}.

\section{Discussion}
\label{sec:discussion}

\begin{figure}
\includegraphics[width=1\columnwidth]{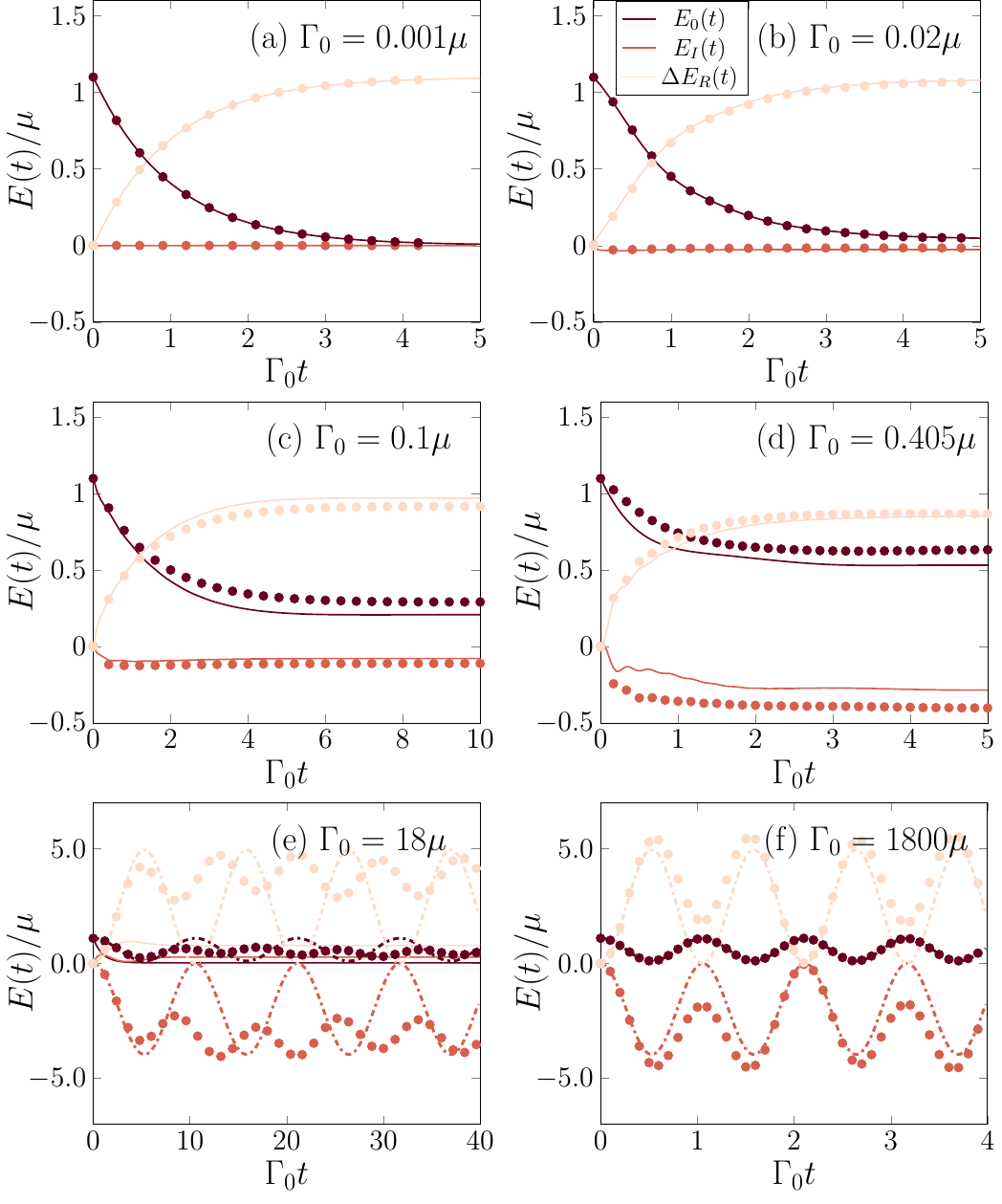}
\caption{Time dependence of the energies $E_0(t)$ and $\Delta E_R(t)$ at weak (a,b), intermediate (c,d), strong coupling (e) and ultra-strong coupling (f). The dots are obtained by numerical evaluation of the exact time evolution \eqref{U}. The full lines show the analytical weak-coupling predictions (\ref{dER}) and the dashed lines the ultra-strong coupling predictions \eqref{ultra:n0} and \eqref{dER_ultra}.
We have assumed $T=0$, $n_0=1$,  $\epsilon_0=1.1\mu$, and $\epsilon_c=10\mu$. For (a) the number of modes in the reservoir is $N=7000$ and the level spacing $\Delta\omega=\epsilon_c/N=1/700\mu$, for (b), $N=500$, $\Delta\omega=1/50\mu$ and for (c-f) $N=200$, $\Delta\omega=1/20\mu$.}
\label{Fig:Et}
\end{figure}

In Fig. \ref{Fig:Et} (a) we illustrate the exponential time dependence of the energies $\langle E_0(t)\rangle$ and $\langle\Delta E_R(t)\rangle$ at zero temperature and at weak coupling. 
We assumed that the resonant level was initially populated, $n_0=1$. 
For the chosen parameters, namely,
$\Gamma_0=0.001\mu$, $\epsilon_0=1.1\mu$, and $\epsilon_c=10\mu$, 
the full solution (\ref{n0},\ref{dER}) and the weak coupling approximation (\ref{dER_weak})
produce overlapping curves. 
The dots in Fig. \ref{Fig:Et} (a) are the result of numerical evaluation of the exact dynamics \eqref{U} as described in Sec. \ref{sec:num}. 
We find that the numerics quantitatively validate the asymptotic analysis, as expected in this parametric range.

In Fig. \ref{Fig:Et} (b)-(f) we show the time dependence of the reservoir energy (\ref{dER}), the energy of the resonant level $\langle E_0(t)\rangle$
and the interaction energy at  $T=0$, with the same values of $\epsilon_0$ and $\epsilon_c$, but at stronger coupling. 
Strong coupling manifests itself in two ways: (i) the interaction energy
$\langle E_I(t)\rangle$ becomes comparable with $\langle \Delta E_R(t)\rangle$ and $\langle E_0(t)\rangle$, 
and (ii) oscillatory contributions to the average energies $\propto e^{-\Gamma_0t/2}\cos[(\tilde\epsilon_0-\mu)t]$ become visible. 
We consider four regimes of system-reservoir coupling 
as discussed in \cite{Wolkanowski2013}: weak ($\Gamma_0=0.02\mu$), interdemiate ($\Gamma_0=0.1\mu$ and $0.405\mu$), strong ($\Gamma_0=18\mu$) and ultra-strong ($\Gamma_0=1800\mu$) coupling coupling. 
For $\Gamma_0=0.02\mu$ the energies still display almost exponential decay and the numerics (dots) 
and analytical predictions (\ref{n0},\ref{dER}) (full lines) agree quite well. 
In the intermediate regime, $\Gamma_0=0.1\mu$ and $0.405\mu$ , deviations from the exponential decay  and the oscillations
become visible. In addition, at short times the analytic results do not agree with the numerics because the condition 
(\ref{condition}) is no longer valid. At strong coupling 
we observe strong oscillations, which do not decay in time 
in qualitative agreement with our asymptotic analysis (\ref{ultra:n0},\ref{dER_ultra}) and results from spectral analysis, see e.g. \cite{jaksic,Cornean}.
Finally, at ultra-strong coupling we again observe strong oscillations with the frequency $\Delta E_I$
in agreement with Eqs. (\ref{ultra:n0},\ref{dER_ultra}).

\begin{figure}
\includegraphics[width=\columnwidth]{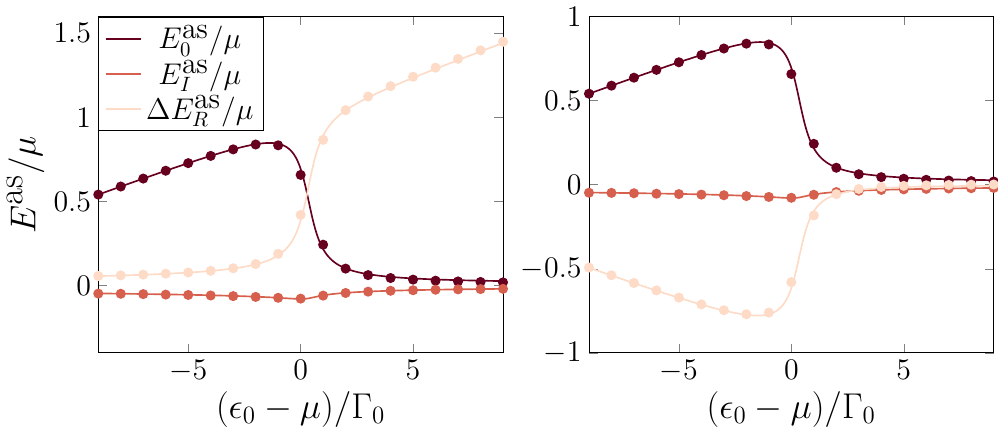}
\caption{Average values of the long time asymptotic energies of the resonant level $\langle E_0^{\rm as}\rangle$ (\ref{E0_eq}), 
of the reservoir $\langle\Delta E_R^{\rm as}\rangle$ (\ref{ER_eq}) and of the interaction energy $\langle E_I^{\rm as}\rangle$ (\ref{EI_eq}) 
plotted versus the energy of the level $\epsilon_0$. The dots are obtained by numerical evaluation of the exact dynamics \eqref{U}. The temperature is zero, $T=0$. 
In the left panel we assume $n_0=1$, i.e. the energy level is initially populated, in the right panel we put $n_0=0$. The coupling rate between the level and the reservoir is $\Gamma_0=0.05\mu$, $\epsilon_c=10\mu$ and $N=200$, $\Delta\omega=1/20\mu$.}
\label{Fig:Eeq}
\end{figure}

Next, we plot the long-time asymptotic energies (\ref{E0_eq}-\ref{EI_eq}) in Fig. \ref{Fig:Eeq}. The asymptotic energy of the resonant level (\ref{E0_eq}) 
is independent on its initial population $n_0$, $\langle E_0^{\rm as}\rangle \approx \tilde\epsilon_0$ for 
$\tilde\epsilon_0<\mu$ and $\langle E_0^{\rm as}\rangle\approx 0$ for 
$\tilde\epsilon_0>\mu$.
In contrast, the asymptotic energy of the reservoir is very sensitive to $n_0$. For $n_0=1$ this energy grows as 
$\langle \Delta E_R^{\rm as}\rangle\approx \epsilon_0$
for $\tilde\epsilon_0>\mu$, since in this case 
the electron escapes from the level to one of the unoccupied states in the reservoir. 
For $\tilde\epsilon_0<\mu$ the electron
stays on the atom since Pauli principle prevents it form jumping to the occupied states in the reservoir,
hence $\langle\Delta E_R^{\rm as} \rangle$ is small. 
With similar arguments, one can easily understand that for  $n_0=0$ 
the reservoir energy should behave as $\langle\Delta E_R^{\rm as}\rangle\approx -\tilde\epsilon_0$ for $\tilde\epsilon_0<\mu$ and 
$\langle \Delta E_R^{\rm as}\rangle\approx 0$ for $\tilde\epsilon_0>0$. In the
vicinity of the Fermi level the switching from one regime to another occurs within the interval $|\tilde\epsilon_0-\mu|\lesssim\Gamma_0$.   
The average interaction energy (\ref{EI_eq}) is always negative, $\langle E_I^{\rm as}\rangle\sim -\Gamma_0$.
In Fig. \ref{Fig:Eeq} we have chosen rather weak coupling, $\Gamma_0=0.05\mu$, therefore the analytical expressions
(\ref{E0_eq}-\ref{EI_eq}) agree with the exact numerics quite well. 

\begin{figure}
    \includegraphics[width=0.8\columnwidth]{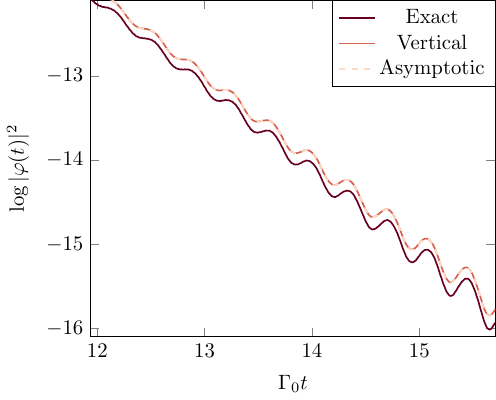}
    \caption{Longtime non-exponential behavior of $|\varphi(t)|^{2}$ for $\Gamma_0=0.0623\,\mu$, $\epsilon_c=10\mu$ and $\epsilon_0=1.1\mu$: three curves, two overlapping.  The dark red curve shows the numerical evaluation of the integral \eqref{phi}. The orange full line and yellow dashed respectively show the numerical evaluation of the vertical contour contribution plus the residue of (\ref{pole:decay}) and asymptotic evaluation (\ref{longtime}) of the vertical contour integral plus residue contribution.
    The small discrepancy between the three curves originates from the residues' contributions. 
    The time scales in the plot require retaining all terms in the analytic asymptotic, as reported in the second row of (\ref{longtime}).}
    \label{fig:logD}
\end{figure}

In Fig. \ref{fig:logD} we show the long time behaviour of the resonant level occupation $|\varphi(t)|^2$ for $t$  longer then the Khalfin time $t_K$.
After a long period of exponential decay, the level enters a regime of power law decay with oscillations.

\begin{figure}
\centering
\includegraphics[width=1\columnwidth]{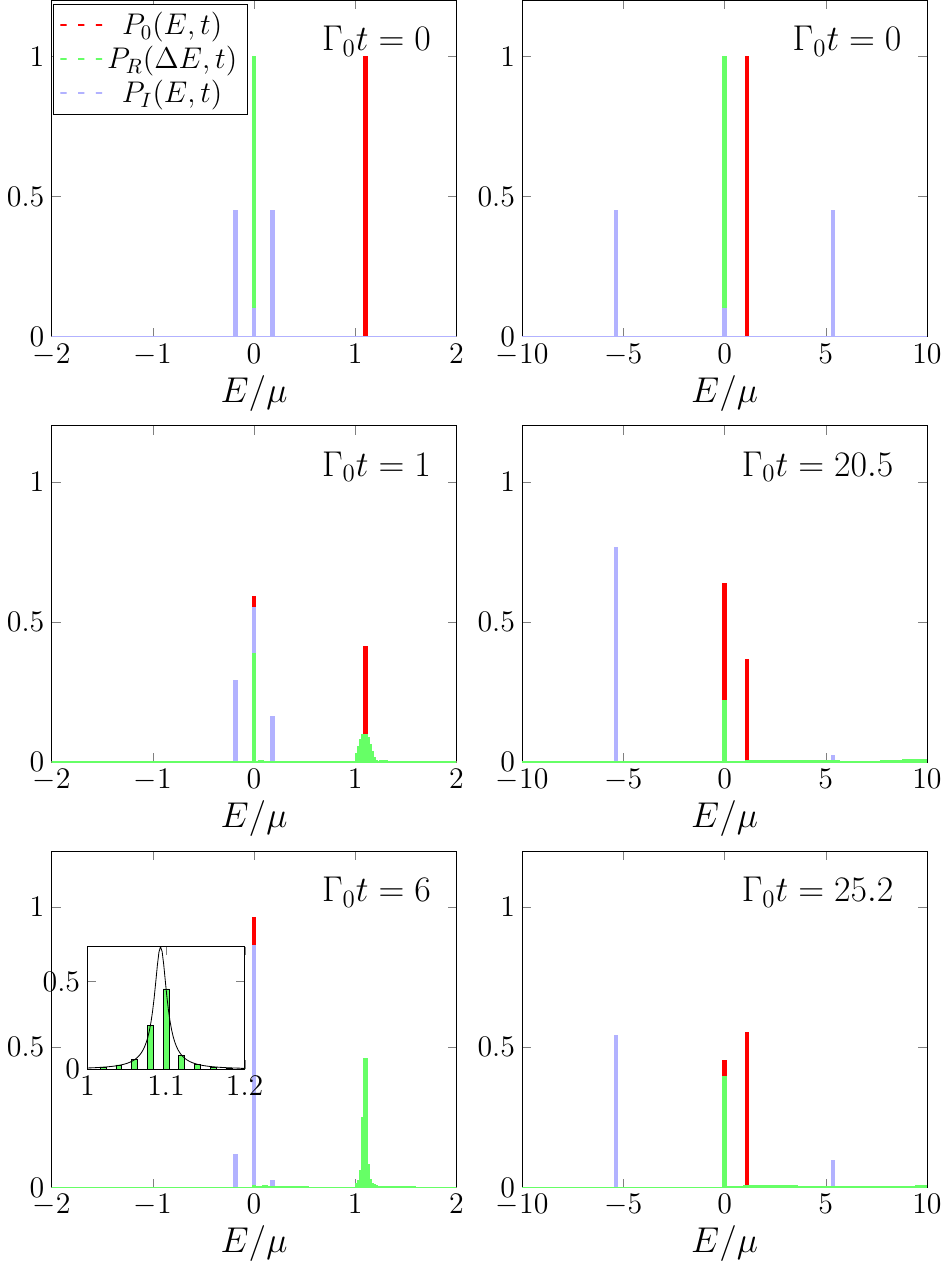}
\caption{Probability distributions of the reservoir, interaction and resonant level energies. Left column corresponds the weak coupling regime with 
$\Gamma_0=0.02\mu$, and the right one -- to the strong coupling limit $\Gamma_0=18$. 
Other parameters have the values $T=0$, $n_0=1$, $\epsilon=1.1\mu$, $\epsilon_c=10\mu$ and $N=200$, $\Delta\omega=1/20\mu$.
The inset in the bottom left panel shows a Lorentzian fit
f the peak in the reservoir energy distribution, $P_R(\Delta E,t)=a/[(\Delta E-\tilde\epsilon_0)^2+\Gamma_0^2/4]$. 
We find from the fit $\Gamma_0=0.0193\mu$, which is close to the expected value $0.02\mu$.}
\label{fig:prob}
\end{figure}

Finally, Fig. \ref{fig:prob} shows the probability distribution of the resonant level,  the reservoir and the interaction energies evaluated
numerically. As expected, the distribution of the level energy (\ref{P0}) has two $\delta$-peaks at $E_0=0$ and $E_0=\epsilon_0$.
Since we have chosen the initial condition $n_0=1$, in the weak coupling regime the distribution of the reservoir energy also has two peaks
centered around $\Delta E_R=0$ and $\Delta E_R=\epsilon_0$. The peak at $\Delta E_R=0$ always remains sharp,
while the second peak at $\Delta E_R=\epsilon_0$ acquires finite length $\sim\Gamma_0$ with growing time.
In fact, one can show that at long times in the weak coupling regime this peak should approach the Lorentzian shape 
$P_R(E,t)\propto \Gamma_0/[(\Delta E_R-\tilde\epsilon_0)^2 + \Gamma_0^2/4]$. Note that in the strong coupling regime the energy distribution of the reservoir does not show a second peak.
Finally, in agreement with the Eq. (\ref{PI}), the distribution of the interaction energy has three sharp peaks separated
by the intervals (\ref{dEI}) $\Delta E_I=\sqrt{\Gamma_0\epsilon_c/2\pi}$.

\subsection{Possible experiment}

In the previous section we have demonstrated that the strong coupling between a resonant energy level
and a metallic reservoir can lead to the non-exponential relaxation of the energy and to significant
part of the energy being stored in the interaction part of the Hamiltonian.  
In this section we will briefly discuss possible experiment, in which these predictions may be tested.
We do not aim at detailed experimental proposal with realistic paramters, rather we limit ourselves
by a qualitative level discussion of possible experimental protocol and relevant time scales in 
nanoelectronic devices.   

\begin{figure}
\includegraphics[width=0.8\columnwidth]{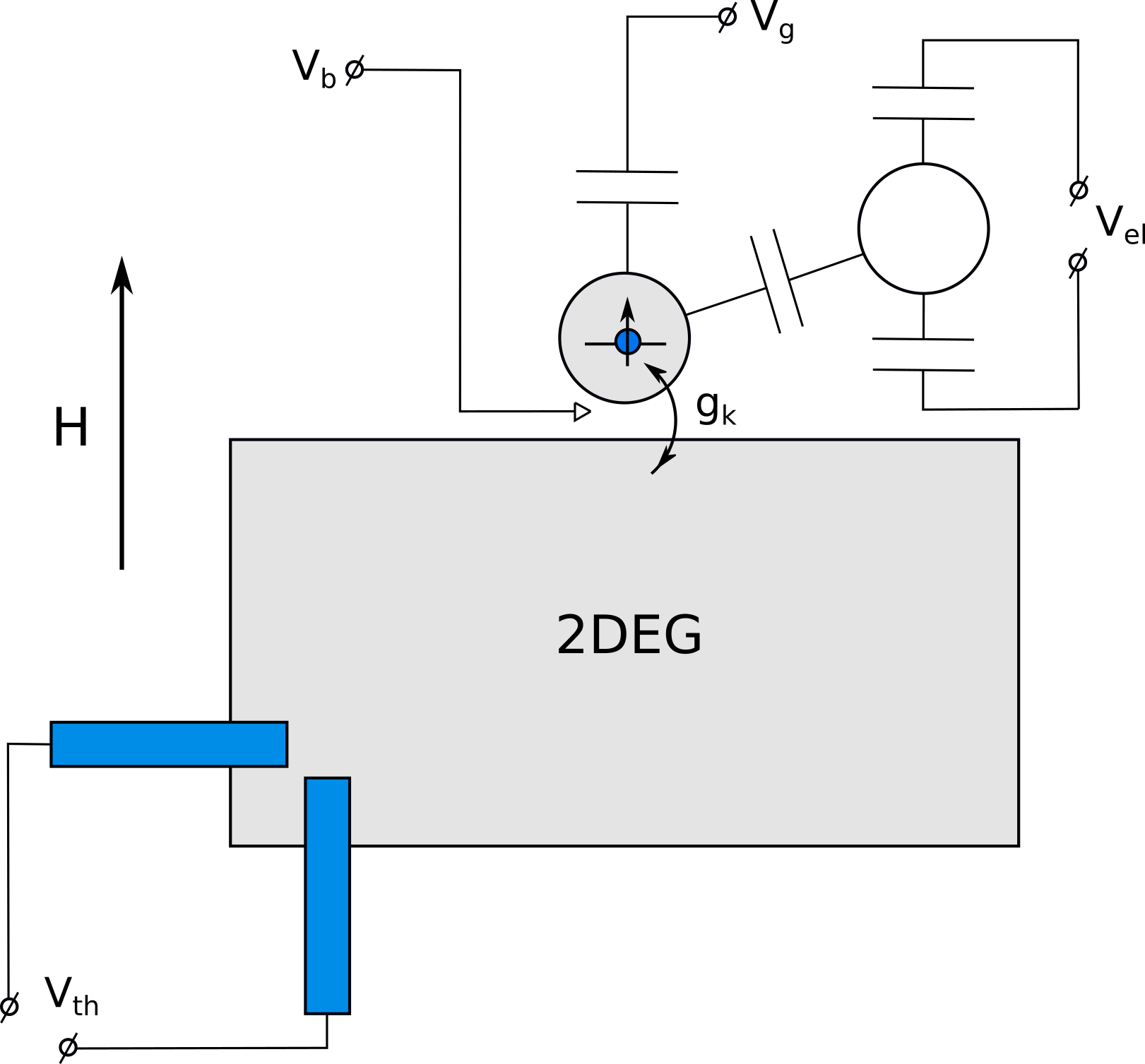}
\caption{Sketch of possible experimental setup. The control potential $V_b$ allows one to tune the height of the barrier 
between the quantum dot hosting a spin polarized energy level and the 2DEG reservoir, 
the gate potential $V_g$ tunes the position of the level relative to the Fermi energy of the reservoir,
the electrometer allows one to monitor the number of electrons in the dot,
and the thermometer in the lower left corner measures the temperature of the reservoir.}
\label{system}
\end{figure}

A possible setup for such an experiment would be a system containing
a finite size area containing a two dimensional electron gas (2DEG), playing the role of the metallic reservoir, 
and a quantum dot with an energy level spin polarized by the strong in plane magnetic field.
This setup is depicted in Fig. \ref{system}. The charging energy of the dot should be small, $E_C\ll\Gamma_0$. 
The barrier between the quantum dot and the reservoir
may be tuned by applying the potential $V_b$ to the control gate electrode; the position of the level
relative to the Fermi energy can be tuned by the gate voltage $V_g$; the number of electrons in the
dot can be detected by an electrometer. Finally, the temperature
of the 2DEG can be monitored, for example, by a thermometer based on superconductor - normal metal -
superconductor Josepshon junction\cite{Libin}.  
The experiment should be run as follows: (i) at time $t=0$ the barrier between the dot and the reservoir
is reduced and they become coupled with the rate $\Gamma_0$, (ii) at final time $t$ this coupling is switched
off again and the reservoir is left to relax, (iii) the number of electrons in the dot is measured, 
(iv) after the delay time, which should be longer than the electron-electron
relaxation time $\tau_{e-e}$, but shorter than the electron-phonon time $\tau_{e-ph}$, the temperature of the 2DEG reservoir is measured.
The measured temperature can be converted to the energy of the reservoir as $\Delta E_R= C_V \Delta T_R$, where
$C_V$ is the heat capacity of the reservoir, and the number of electrons in the dot can be converted into the energy
of the resonant level. Corresponding energy distributions can be obtained by repeating this experiment many times.
The interaction energy cannot be directly measured, but it should be possible to infer its average value from
the energy conservation condition (\ref{conservation}).    
As for the original motivation of our study, one should be able to easily determine the initial population of the quantum dot
level by measuring the temperature of the reservoir if the energy of the level $\epsilon_0$ is sufficiently high. 
Easily achievable values $\epsilon_0 > 100$ $\mu$eV should be sufficient for that.

This type of experiment is certainly challenging because of the short values of the electron-electron relaxation time $\tau_{e-e}$.
Indeed, $\tau_{e-e}$ lies in the nanosecond range at the lowest accessible temperatures\cite{Pothier,Niimi}. This
leaves little room for observation of non-exponential time relaxation and long-time asymptotics (\ref{longtime}).
We believe, however, that one should be able to measure the asymptotic values of the average level and reservoir energies (\ref{E0_eq},\ref{ER_eq}),
and subsequently estimate the interaction energy (\ref{EI_eq}) from the conservation condition (\ref{conservation}). 
The interaction energy should be observable because
the coupling rate can be easily made rather big, $\Gamma_0\gtrsim 100$ $\mu$eV.

In conclusion, we have considered an exactly solvable model of a resonance level coupled
to a metallic reservoir. We have considered the transient process in which the level and the reservoir are coupled at time $t=0$ 
and determined the time dependence of the average values of the reservoir energy, of the resonant level energy and of the interaction
energy in the strong coupling regime. We have also found the statistical distributions of these energies.

\section{Acknowledgement}

We are glad to acknowledge very useful discussions with Jukka Pekola. B.D. was supported by DOMAST.
D.G. was supported by the Academy of Finland Centre of Excellence
program (project 312057).

\end{document}